\documentclass[aps,prb,twocolumn,groupedaddress,showpacs,a4paper]{revtex4}
\usepackage{graphicx}

\begin{document}


\title{Nonlinear dynamics of quantum dot nuclear spins}


\author{P. Maletinsky}
\email[]{patrickm@phys.ethz.ch}
\author{C. W. Lai}
\author{A. Badolato}
\author{A. Imamoglu}
\affiliation{Institute of Quantum Electronics, ETH-H\"{o}nggerberg,
CH-8093, Z\"{u}rich, Switzerland}

\date{\today}

\begin{abstract}
We report manifestly nonlinear dependence of quantum dot nuclear
spin polarization on applied magnetic fields. Resonant absorption
and emission of circularly polarized radiation pumps the resident
quantum dot electron spin, which in turn leads to nuclear spin
polarization due to hyperfine interaction. We observe that the
resulting Overhauser field exhibits hysteresis as a function of
the external magnetic field. This hysteresis is a consequence of
the feedback of the Overhauser field on the nuclear spin cooling
rate. A semi-classical model describing the coupled nuclear and
electron spin dynamics successfully explains the observed
hysteresis but leaves open questions for the low field behaviour
of the nuclear spin polarization.
\end{abstract}


\maketitle


\section{Introduction}

Coupling of a single confined electron-spin to a mesoscopic
ensemble of nuclear spins defined by a quantum dot (QD) gives rise
to rich physical phenomena such as non-Markovian electron-spin
decoherence.\cite{Khaetskii2002,Johnson2005,Koppens2005} It has
been argued that controlling QD nuclear spins by carrying out
either dynamical nuclear spin polarization (DNSP) or precise
measurements of the Overhauser field would prolong the electron
spin coherence time and thereby enhance the prospects of
implementing QD spin-based quantum information
processing.\cite{Coish2004} Several groups have previously
reported QD nuclear-spin cooling in various QD
systems.\cite{Gammon1997,Ono2004}A non-trivial dependence of DNSP
on the applied external magnetic field however, was never observed
at high fields. Recently, DNSP in self-assembled InGaAs QDs was
reported for low external magnetic fields.\cite{Eble2005,Lai2006}
Experiments carried out using resonant excitation of the QD
excited-states showed that DNSP can be observed even when the
external field vanishes: this has been attributed to the role of
the inhomogeneous electronic Knight field in suppressing
depolarization due to the spin-non-conserving part of the nuclear
dipole-dipole interactions. An obvious extension of this work is
to study the limits on the degree of DNSP that can be attained in
these QDs by, for example, investigating the external magnetic
field dependence.

In this work, we study the magnetic field dependence of DNSP in a
single self-assembled QD. The QD is located in a gated structure
that allows for deterministic QD charging with a single excess
electron\cite{Warburton2000} or hole.\cite{Edinger2005} The
photoluminescence (PL) polarization and spin splitting are studied
by resonantly exciting the QD in one of its (discrete) excited
($p$-shell) states under external magnetic fields ($B_{\rm ext}$)
ranging from $B_{\rm ext}=-2$ to $2~$T, applied along the crystal
growth $z$-axis at $T=2.5~$K.
The PL spectral lines associated with different charging states of
a single QD can be identified from a PL intensity contour plot as
a function of the bias voltage and emission
energy.\cite{Warburton2000} In this work, we focus on the spectral
features of the negatively charged exciton ($X^{-1}$), consisting
of two electrons in a spin singlet state and one hole trapped in
the QD. This charge complex has been shown to lead to a
polarization of the underlying nuclear spin system under
circularly polarized excitation.\cite{Lai2006}

The sample was grown by molecular beam epitaxy on a $(100)$
semi-insulating GaAs substrate. The InAs QDs are spaced by $25~$nm
of GaAs from a $40~$nm doped n$^{++}$-GaAs layer, followed by
$30~$nm GaAs and 29 periods of AlAs/GaAs ($2/2~$nm) superlattice
barrier layer, and capped by $4$-nm GaAs. A bias voltage is
applied between the top Schottky and back ohmic contacts to
control the charging state of the QDs. The low density of QDs
($<0.1 \: \mu $m$^{-2}$) allows us to address a single QD using a
micro-photoluminescence ($\mu$-PL) setup.

Our standard $\mu$-PL setup\cite{Lai2006} is based on the
combination of a solid immersion lens, directly fixed onto the
sample, and a focussing lens mounted outside the cryostat. The
sample is placed in a Helium-bath cryostat equipped with a
superconducting magnet, reaching a maximum magnetic field strength
of $10~$T and oriented in the Faraday geometry. The spectroscopy
system consists of a spectrometer of $0.75~$m focal length and a
liquid-nitrogen cooled CCD camera providing a spectral resolution
of $\sim20~\mu$eV. The energies of the QD PL lines are determined
by estimating the center of mass of the observed emission lines by
calculating a weighted average over the relevant CCD pixels. We
estimate the resulting accuracy in our energy measurement to be
$\sim2~\mu$eV.



We have performed measurements on the negatively charged exciton
$X^{-1}$ at the center of its PL stability region with respect to
gate voltage. In this regime, electron co-tunnelling to the nearby
reservoir has shown to be minimized\cite{Atature2006} and the QD
is occupied with a single electron in its ground state. Optical
excitation is performed in a resonant way into the $p$-shell,
which lies approximately one LO phonon energy above the emission
energy of $X^{-1}$ ($E_0=1.316524~$eV). The excitation power is
fixed close to saturation of the observed emission line. We found
that these conditions lead to a maximal preservation of PL light
polarization
($\sim75\%$ at $B=0~$T) after excitation with circularly polarized
light.

For $X^{-1}$, circular polarization of the emitted light reflects
both the spin of the hole in the QD state before emission and the
initial spin of the residual electron in the QD after photon
emission. A high degree of circular polarization of
$X^{-1}$-emission thus indicates a highly spin polarized residual
electron in the QD. Since the electron spin system is in thermal
contact with the nuclear spin system through the hyperfine
interaction, spin polarization (or temperature) will be
transferred from one to the other, thereby cooling the nuclear
spin system. At the same time, nuclear spin diffusion, quadrupolar
interactions and other nuclear spin relaxation mechanisms will
heat up the nuclear spin system, leading to its finite spin
temperature in a dynamical equilibrium. In the following, we want
to study these spin-transfer mechanisms and the dependence of the
final nuclear spin temperature on an external magnetic field
$B_{\rm ext}$ along the spin quantization axis.

\section{Hyperfine Interaction}
The dominant contribution to the coupling between the electron-
and the nuclear-spin systems originates from the Fermi contact
hyperfine interaction. For an electron in a QD and in first order
perturbation theory, this can be written as\cite{Merkulov2002}
\begin{equation}\label{Hyperfine}
\hat{H}_{\rm hf} = \frac{\nu_0}{8} \sum_i A_i
|\psi(\textbf{R}_i)|^2 \hat{\textbf S} \cdot \hat{\textbf I}^i,
\end{equation}
where $\nu_0$ is the volume of the InAs-crystal unit cell
containing eight nuclei, $\hat{\textbf S}$ is the dimensionless
electron spin operator, $\psi(\textbf{r})$ is the electron
envelope wave function and $\hat{\textbf I}^i$ and $\textbf{R}_i$
are the spin and location of the i-th nucleus, respectively.
$A_i=(2\mu_0 g_0 \mu_{\rm B}\mu_i/3I^i)|u(\textbf{R}_i)|^2$ is the
hyperfine coupling constant, which depends on the nuclear magnetic
moment $\mu_i$, the nuclear spin $I^i$ and on the value of the
electron Bloch function $u(\textbf{R}_i)$ at the nuclear site.
$\mu_{\rm B}$ is the Bohr magneton, $g_0$ the free electron
g-factor and $\mu_0$ the permeability of free space. $A_i$ is
positive and is on the order of $50~\mu$eV for all the nuclei in
our system.

With the identity $\hat{\textbf S} \cdot \hat{\textbf
I}^i=1/2(I^i_+S_-+I^i_-S_+)+I^i_zS_z$ where $I^i_{\pm}$ and
$S_{\pm}$ are the nuclear and electron spin raising and lowering
operators respectively, equation (\ref{Hyperfine}) can be
decomposed into two parts:\cite{Meier1984} A dynamical part
($\propto I^i_+S_-+I^i_-S_+$), allowing for the transfer of
angular momentum between the two spin systems and a static part
($\propto I^i_zS_z$), affecting the energies of the two spin
systems. In the absence of any other relaxation mechanisms, the
dynamical contribution leads to an equilibrium mean nuclear spin
polarization $\langle I_z^i \rangle$ along the quantization axis
$z$, given by \cite{Abragam1961}
\begin{equation}\label{EquilibriumI}
\langle I_z^i \rangle=\frac{I^i(I^i+1)}{S(S+1)} \langle S_z
\rangle,
\end{equation}
where $\langle S_z \rangle$ is the mean electronic spin along the
$z$-axis. This equation is valid if $\langle I_z^i \rangle\ll I^i$
and if we neglect any polarization due to thermalization in an
external magnetic field of either of the two spin systems.

The static part leads to the notion of the ``effective magnetic
field'', either seen by the electron due to spin polarized nuclei
(Overhauser field), or by the nuclei due to a spin polarized
electron (Knight field). The effects of the Knight field on the
order of $10-100~$G upon the nuclear spin system have been studied
elsewhere \cite{Lai2006} and will not be considered in the present
work. The Overhauser field operator can be written as
\begin{equation}\label{Overhauserfield}
\hat{B}_{\rm nuc} = \frac{1}{g_{\rm el}^*\mu_{\rm
B}}\frac{\nu_0}{8} \sum_i A_i |\psi(\textbf{R}_i)|^2 \hat{I_z^i},
\end{equation}
and has a finite expectation value $B_{\rm nuc}$ if the nuclei are
partly polarized. This effective field leads to a total electron
Zeeman splitting in the presence of both, nuclear and external
magnetic fields of
\begin{equation}\label{DeEnergy}
\Delta E^Z_{\rm el}=g_{\rm el}^* \mu_{\rm B} (B_{\rm ext}+B_{\rm
nuc}).
\end{equation}
The energy shift due to spin polarized nuclei is referred to as
Overhauser shift (OS). We note that only electrons in the
conduction band experience a substantial OS. For carriers in the
valence band, the contact hyperfine interaction (\ref{Hyperfine})
vanishes due to the p-type symmetry of this band in \textit{III-V}
semiconductors. Since the electrons in $X^{-1}$ are in a singlet
state, they are not affected by $B_{\rm nuc}$ either and only the
final state of recombination shifts due to nuclear polarization.
The total Zeeman splitting of the $X^{-1}$ recombination line thus
amounts to
\begin{equation}\label{DXEnergy}
\Delta E^{\rm Z}_{X^{-1}}=-g_{\rm h}^* \mu_{\rm B} B_{\rm
ext}-g_{\rm el}^* \mu_{\rm B} (B_{\rm ext}+B_{\rm nuc}),
\end{equation}
where $g_{\rm el}^*$ and $g_h^*$ are the electron- and hole
g-factors, respectively.

Exciting the QD with linearly polarized light creates residual
electrons in a superposition of spin up and down, resulting in no
nuclear polarization and $B_{\rm nuc}=0~$T. Thus, comparing the
Zeeman splittings of $X^{-1}$ under linearly- and circularly
polarized excitation ($\Delta E^{\rm Z,lin.}_{X^{-1}}$ and $\Delta
E^{\rm Z,\sigma^\pm}_{X^{-1}}$, respectively) gives a direct
measure of $B_{\rm nuc}$:
\begin{equation}\label{DZEnergy}
\Delta E^{\rm Z,\sigma^\pm}_{X^{-1}}-\Delta E^{\rm Z,
lin.}_{X^{-1}} = - g_{\rm el}^* \mu_{\rm B} B_{\rm nuc}.
\end{equation}

\section{Experimental procedure}

\begin{figure}[t]
\includegraphics[width=\columnwidth]{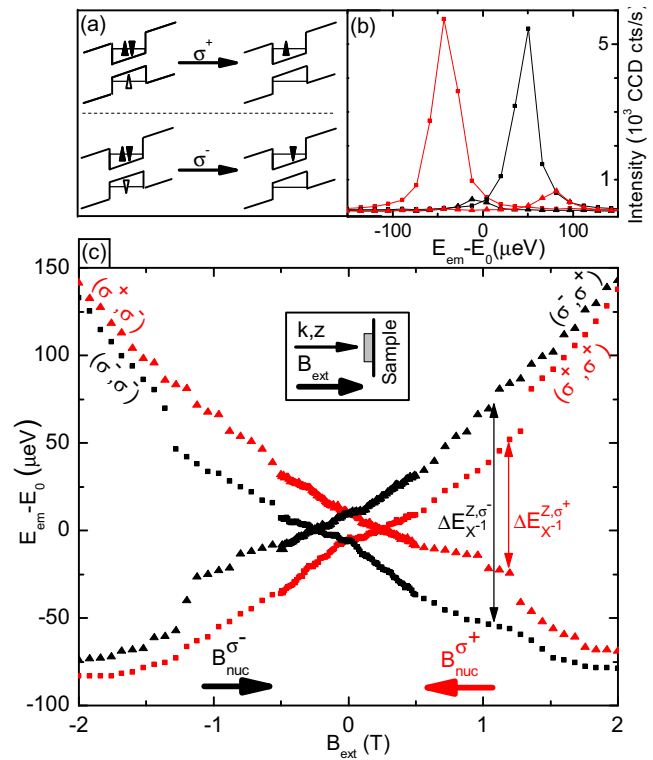}
\caption{\label{FigLineshifts}(color online). \textbf{(a)} Spin
configurations of $X^{-1}$ before and after the emission of a
$\sigma^\pm$ polarized photon. Open (filled) triangles denote the
spin of the hole (electron). \textbf{(b)} Raw spectra at $B_{\rm
ext}=-0.96~$T for the four excitation/detection configurations in
the circular basis: red (black) denotes excitation with $\sigma^+$
($\sigma^-$) polarized light. Detection is co- or cross-circular
(squares and triangles, respectively). \textbf{(c) Energy
dispersion of $X^{-1}$ under circularly polarized excitation:} The
different emission energies ($E_{\rm em}$) for the two excitation
polarizations are due to dynamical nuclear spin polarization which
leads to an effective nuclear magnetic field $B_{\rm
nuc}^{\sigma\pm}$ under $\sigma^{\pm}$ excitation (orientation
indicated by the arrows in the figure). The energy
$E_0=1.316524~$eV of the $X^{-1}$ emission under linearly
polarized excitation at $B=0~$T is subtracted from the data.
The inset shows the relative orientation of $k$-vector,
quantization axis $z$ and positive magnetic field.}
\end{figure}

In order to gain knowledge about the nuclear polarization and its
dependence on an external magnetic field along the spin
quantization axis in QDs, we performed a $\mu$-PL experiment on a
single QD in an external magnetic field with the experimental
procedure proposed above. Nuclear polarization manifests itself in
a difference in emission energies between excitation with
circularly and linearly polarized light as was established in the
previous section. Figure~\ref{FigLineshifts} shows the
$X^{-1}$-emission energies of a single QD under excitation with
circularly polarized light as a function of external magnetic
field. Throughout this paper, red (black) color denotes excitation
with $\sigma^+$ ($\sigma^-$) light, while squares (triangles)
stand for co- (cross-) circular detection. The polarizations for
excitation and detection are denoted as
$(\sigma^\alpha,\sigma^\beta)$ where $\sigma^\alpha$ and
$\sigma^\beta$ correspond to excitation and detection,
respectively. The index $\alpha$ or $\beta$
($\alpha,\beta\in[+,-]$) stands for circularly polarized light
with positive or negative helicity. The data shown in figure
\ref{FigLineshifts} was obtained in a single sweep from $B=-2~$T
to $B=+2~$T, varying excitation and detection polarization for
each B-field value in the order
$(\sigma^+,\sigma^-)\Rightarrow(\sigma^+,\sigma^+)\Rightarrow(\sigma^-,\sigma^+)\Rightarrow(\sigma^-,\sigma^-)$
such that any memory of the nuclear spin system is erased during
the sweep. The data for $|B|<500~$mT was taken
with smaller magnetic field steps in order to highlight the
detailed behavior of DNSP at low fields. Every data point
represents the center of mass of the emission peak of $X^{-1}$
taken from a single spectrum with $1~$s integration time and a
signal to noise ratio of $\sim100:1$ for co-circular detection.
The effects of nuclear polarization can be seen in the range of
$|B_{\rm ext}|<1.2~$T where emission energies for a given
detection polarization depend strongly on the helicity of the
laser light. Excitation with $\sigma^+$ light creates a residual
electron with its spin pointing in the positive $z$-direction (see
Fig.\ref{FigLineshifts}). According to equations (\ref{Hyperfine})
and (\ref{Overhauserfield}), this creates a nuclear spin
polarization in the same direction and, due to the negative sign
of the $g_{\rm el}^*$, a nuclear field pointing in the negative
$z$-direction. This scenario is consistent with the polarization
sequences and lineshifts observed in Fig.~\ref{FigLineshifts}.
Above $1.2~$T, the emission energies of the QD are almost
independent of excitation light polarization, indicating that
nuclear effects become very small. Another striking feature in
this data is the symmetry under simultaneous reversal of the
excitation light helicity and the sign of the magnetic field.
However, the data is not symmetric under the reversal of only one
of these parameters. This asymmetry indicates that the system
distinguishes between nuclear fields pointing along or against the
external magnetic field - we will see in the following that it is
more efficient for the system to create a nuclear field pointing
against $B_{\rm ext}$ than one that points along this field.

\begin{figure}[t]
\includegraphics[width=\columnwidth]{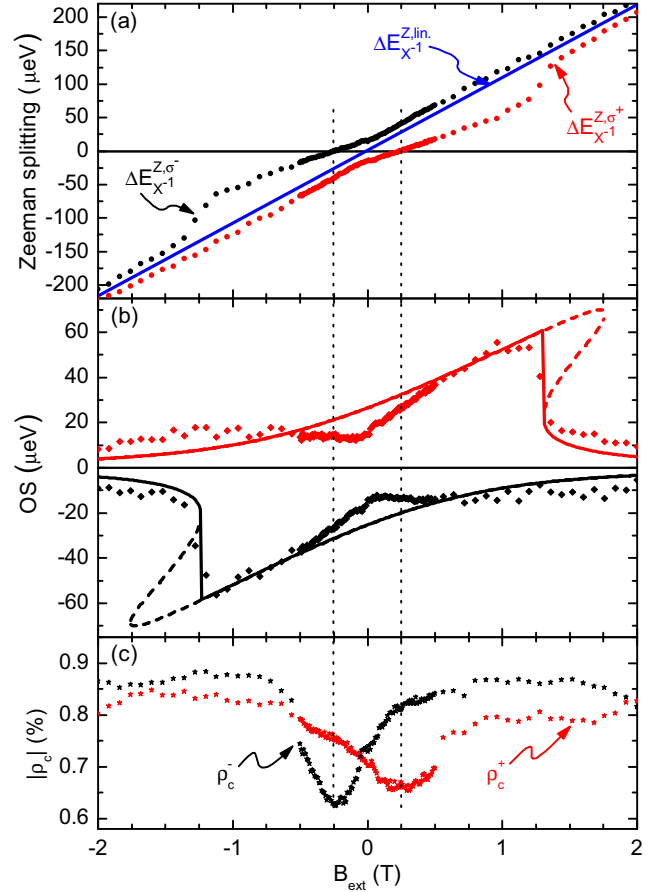}
\caption{\label{FigZeemanOSPol}(color online). \textbf{Nuclear
Polarization in external magnetic fields:} \textbf{(a)} Spin
splitting of $X^{-1}$ under circularly polarized excitation
(Circles). Red represents $\sigma^+$, black $\sigma^-$ excitation.
The solid line is a linear fit to the data as described in the
text. \textbf{(b)} Deviation of spin splitting between circular
and linear excitation: Overhauser shift for $\sigma^+$ and
$\sigma^-$ excitation (red and black diamonds, respectively). The
solid and dashed lines are the results of the fits according to
the model discussed in the text. \textbf{(c)} QD spin polarization
$\rho_c^\pm$ of PL light under $\sigma^\pm$ excitation. The
polarization shows a minimum at the magnetic field where the
Zeeman splitting is zero, consistent with our model of carrier
relaxation.\cite{ Lai2006}}
\end{figure}

In order to obtain a more quantitative picture of the magnetic
field dependent DNSP, we performed the following analysis steps on
the data (see Fig.~\ref{FigZeemanOSPol}): We first extract the
Zeeman splittings for excitation with $\sigma^+$ and $\sigma^-$
light from the raw data shown in Fig.~\ref{FigLineshifts}. To this
data, we fit a linear Zeeman splitting such that the fit coincides
with the data at magnetic fields $B_{\rm ext}>1.8~$T where nuclear
polarization is very small (Fig.~\ref{FigZeemanOSPol}(a)). The
excitonic g-factor, $g_{\rm ex}=1.87$\footnote{We use the
following sign convention of Zeeman Hamiltonians for electrons,
holes and excitons, respectively: $H_{\rm el}= \mu_B g_{\rm el}^*
\textbf{S}_{\rm el}^{} \cdot \textbf{B}$, $H_{\rm h}= - \mu_B
g_{\rm h}^* \textbf{S}_{\rm h}^{} \cdot\textbf{B}$, $H_{\rm
ex}=\frac{1}{2}\mu_B g_{\rm ex} \textbf{S}_{\rm
ex}\cdot\textbf{B}$. In this representation, the heavy-hole wave
functions $|\pm\frac{3}{2}\rangle$ convert to pseudo-spins
$|\pm\frac{1}{2}\rangle$. With this convention, the exciton
g-factor amounts to $g_{\rm ex}^{}=-g_{\rm el}^*-g_{\rm h}^*$.
Experimentally, we then find $g_{\rm el}$ and $g_{\rm h}$ to be
$<0$.}, that we find with this fitting procedure matches within a
few percent to an independent measurement of $g_{\rm ex}$ that we
performed with linearly polarized excitation (not shown here). The
Overhauser shift can now be extracted from this fit with the help
of equation (\ref{DZEnergy}); the result is plotted in
figure~\ref{FigZeemanOSPol}(b). There's a striking difference when
polarizing the nuclei along or against the external field: Nuclear
polarization with $B_{\rm nuc}$ pointing along the applied field
is rather inefficient and shows a slight decrease with increasing
magnitude of the applied field. Polarization with $B_{\rm nuc}$
pointing against the external magnetic field on the other hand
shows a much richer behavior: The nuclear polarization first
increases almost linearly as the magnitude of the external field
increases and then shows a sudden drop when $|B|>1.2~$T.

From the spectral data we can also extract information about the
hole spin polarization before- and the residual electron spin
polarization after recombination of $X^{-1}$. For this, we define
a degree of QD spin polarization as
$\rho_c^\pm:=(I_{(\sigma^\pm,\sigma^+)}-I_{(\sigma^\pm,\sigma^-)})/(I_{(\sigma^\pm,\sigma^+)}+I_{(\sigma^\pm,\sigma^-)})$
under $\sigma^\pm$ excitation. $I_{(\sigma^\alpha,\sigma^\beta)}$
are the intensities of the dominant PL-peaks in the corresponding
analyzer/polarizer configurations ($\alpha,\beta\in[+,-]$). We
note that at zero magnetic field, $\rho_c^\pm$ is identical to the
degree of circular polarization of the single PL peak observed.
The measured quantity $\rho_c^\pm$ is plotted in
figure~\ref{FigZeemanOSPol}(c) as a function of external magnetic
field. It is roughly constant and on the order of $85\%$ over a
wide range of magnetic fields. Only for the fields where the trion
Zeeman splitting vanishes, $\rho_c^\pm$ shows a dip to roughly
$65\%$. This behavior is consistent with the rotation of the
exciton spin during relaxation of the optically created electron
from the excited p-shell state to the s-shell via the electron
reservoir.\cite{Lai2006} During the relaxation, the QD is left
neutral and anisotropic exchange interaction will rotate the
exciton spin. This rotation is most efficient in the absence of
excitonic Zeeman splitting which explains the magnetic field
dependence of PL polarization observed in this measurement.

We note however that there is a certain asymmetry in the data
shown in figure~\ref{FigZeemanOSPol}(c) that remains unexplained:
$\rho_c^-$ is larger than $\rho_c^+$ at high magnetic fields and
the dip in $\rho_c^-$ at lower fields is less pronounced than for
$\rho_c^+$. A possible reason for this asymmetry could be the
different excitation efficiencies in the QD for $\sigma^+$ and
$\sigma^-$ excitation.

\section{Modelling of the data}

Most of the above-mentioned nuclear effects in the presence of an
external magnetic field can be described by a simple rate equation
model already proposed earlier\cite{Gammon2001,Eble2005} and
originally based on the work by Abragam.\cite{Abragam1961}
The rate equation is based on the condition for thermal
equilibrium (\ref{EquilibriumI}) between the electron and the
nuclear spin system in the absence of any coupling to the
environment. This equilibrium is then reached on a typical
timescale given by the nuclear spin relaxation time $T_{\rm 1e}$,
which can be estimated to be\cite{Meier1984}
\begin{equation}\label{T1e}
\frac{1}{T_{\rm 1e}}=
\frac{1}{T_{\rm 1e}^0}\frac{1}{1+\Omega_{\rm el}^2\tau_{\rm
el}^2}.
\end{equation}
Here, $\tau_{\rm el}$ is the electron spin correlation time which
broadens the electronic spin states. $\Omega_{\rm el}=\Delta
E^Z_{\rm el}/\hbar$ is the electron Larmor frequency. This
frequency itself depends on the degree of nuclear polarization
through (\ref{DeEnergy}) and (\ref{Overhauserfield}). For a given
nuclear species, the nuclear spin relaxation time at zero electron
Zeeman splitting is given by $1/T_{\rm 1e}^0= f_{\rm el} \tau_{\rm
el} A_i^2/(N \hbar)^2$ with $N$ the number of relevant nuclei and
$f_{\rm el}$ the fraction of time the QD is occupied with a single
electron. This expression for $T_{\rm 1e}^0$ is valid if we assume
a homogenous electron wave function
$\psi(\textbf{r})\propto\sqrt{8/\nu_0 N}$ which is constant within
the QD volume and zero outside.

By adding a nuclear spin decay channel which is dominated by
nuclear spin diffusion out of the QD on a timescale $T_{\rm d}$,
we end up with a rate equation of the form
\begin{equation}\label{Rateequation}
\frac{d\langle I_z^i \rangle}{dt} = - \frac{1}{T_{\rm 1e}}(\langle
I_z^i \rangle-\frac{4}{3}I^i(I^i+1)\langle S_z
\rangle)-\frac{1}{T_{\rm d}}\langle I_z^i \rangle.
\end{equation}
This equation was obtained for the coupling of a single electron
to a single nuclear spin. It can be approximately generalized to
the case of an ensemble of different nuclei in the QD by
considering the mean nuclear spin polarization $\langle I_z
\rangle=\frac{1}{N} \sum_i \langle I_z^i \rangle$. For this, we
replace the hyperfine constant $A_i$ in (\ref{T1e}) and the
quantity $I^i(I^i+1)$ in (\ref{Rateequation}) each by a weighted
average $\overline{\zeta}=0.5(x \zeta_{In}+(x-1)
\zeta_{Ga}+\zeta_{As})$. $x$ and $(1-x)$ are the relative contents
of In and Ga in the QD (a typical value for our QDs is $x=0.9$)
and $\zeta$ represents the variable to be averaged over the
different nuclear species ($A_i$ or $I^i(I^i+1)$).
With values of $A_i$ taken from the
literature\cite{Chemistry1968}, this results in
$\overline{A^i}=50.3~\mu$eV and $\overline{I^i(I^i+1)}=13.2~$. We
take these numbers to be fixed in the following, even though in
reality they might vary due to uncertainties in QD composition and
confined electron wave-function.

Since the electron-mediated nuclear spin relaxation time $T_{\rm
1e}$ itself depends on nuclear spin polarization, equation
(\ref{Rateequation}) leads to the following self-consistent
nonlinear steady state solution $\langle I^{\rm ss}_z \rangle$ for
the mean nuclear spin polarization:
\begin{equation}\label{IzSteadySt}
\langle I^{\rm ss}_z
\rangle=\frac{4}{3}\frac{\overline{I^i(I^i+1)} \langle S_z
\rangle}{1+ \frac{T_{\rm 1e}^0}{T_{\rm d}}(1+(\frac{\tau_{\rm
el}}{\hbar})^2(g_{\rm el}^*\mu_{\rm B} B+\overline{A_i}\langle
I^{\rm ss}_z \rangle)^2)}.
\end{equation}
For the fitting procedure we take $\langle S_z \rangle$ to be
independent of magnetic field and equal to half the mean PL
polarization $\rho_c^\pm/2$ observed in the experiment for
$|B_{\rm ext}|>0.5~$T. We note that using the magnetic field
dependent $\rho_c^\pm$ measured in the experiment
(Fig.~\ref{FigZeemanOSPol}(c)~) did not lead to a significant
improvement of the fits and is thus not shown here. It also has to
be noted that the way we averaged equation (\ref{IzSteadySt}) over
the different nuclear species as well as the fact that we used a
homogenous wavefunction for the electron and that we neglected the
magnetic field dependence of $T_{\rm d}$ all limit the validity of
this model.

We numerically solved the implicit equation (\ref{IzSteadySt}) in
order to fit the data. The result of such a numerical solution is
shown in figure~\ref{FigZeemanOSPol}(b). The model qualitatively
reproduces the data. Still,  some features, like the fast change
of DNSP around zero external field as well as the high residual
spin polarization at high external magnetic fields, could not be
explained within the model. In the region $1.2~$T$<B_{\rm
ext}<1.8~$T the model predicts three solutions: two stable states,
one with a low and one with a high degree of DNSP and an unstable
solution of intermediate nuclear spin polarization (the last two
solutions correspond to the dashed lines in
Fig.~\ref{FigZeemanOSPol}). Since in this experiment we changed
excitation polarization from $\sigma^+$ to $\sigma^-$ for each
magnetic field value, the system always followed the solution with
minimal nuclear spin polarization and DNSP dropped at $B_{\rm
ext}\simeq1.2~$T. The fact that the drop in DNSP in this
measurement was rather smooth compared to the model prediction was
probably due to the very long timescale of the buildup of DNSP
right before its disappearance: since in the experiment every
point was taken with an integration time of $1~$s, the nuclear
system did not have time to reach its steady state polarization
before the excitation light polarization was switched. We will
discuss this breakdown of DNSP as well as the regime of high
nuclear spin polarization and bistability in more detail in the
following section. The parameters used for the fitting curve in
figure~\ref{FigZeemanOSPol}(b) were
$T_{\rm 1e}^0/T_{\rm d}=4.3$,$\rho_c=0.84$,$\tau_{\rm el}=35~$ps,
$g_{\rm el}^*=-0.69$, which are all realistic values for our QD.
The electron spin correlation time found in the fit can be
explained in a simple three level picture where the QD is excited
from its ground state into its $p$-shell and PL emission is
observed from carriers recombining from the $s$-shell. Since this
system is pumped close to saturation, the lifetime and thus the
coherence time of the residual electron are limited by the
relaxation time from the $p$-shell to the n$^{++}$-GaAs layer by
tunnelling. This timescale is expected to be shorter than
$20~$ps.\cite{ Lai2006}

The parameters obtained in this fit also allow us to estimate the
nuclear spin relaxation time $T_{\rm 1e}^0$. Using the value
$\tau_{\rm el}=35~$ps, the corresponding value for $f_{\rm
el}=0.035$ (assuming an exciton lifetime of $1~$ns) and
N$=10^4-10^5$, we obtain $T_{\rm 1e}^0=0.1-1~$s. This value is
roughly consistent with the buildup time of nuclear spin
polarization we observed.\cite{ Lai2006}

We extended the presented model by including the dynamics of the
mean electron spin $\langle S_z \rangle$. This leads to a rate
equation for the electron spin of a form similar to equation
(\ref{Rateequation}).
The main differences between the electron and the
nuclear spin dynamics are that the electron
spin system in the absence of losses reaches the thermal
equilibrium state (\ref{EquilibriumI}) at a rate $N/T_{\rm 1e}$.
Compared to the nuclear spin relaxation rate, the electron spin
relaxation is faster by the number of nuclei $N$ in the system. In
addition, the electron spin is repumped into its initial state
$S_z^0=\rho_c^\pm/2$ at the $s$-shell decay rate on the order of
$1~$ns. This extension however, did not lead to any new insights
on the behavior of the nuclear spin system. A numerical study of
this extended model suggested though that the mean electron spin
decreases linearly with increasing nuclear spin polarization. The
electron spin thus seems to follow the intricate dynamics of the
nuclear spin system. This observation motivates further studies on
the positively charged exciton where PL light polarization gives a
direct measure of the mean electron spin.


\section{Hysteresis in the magnetic field sweeps}

\begin{figure}[t]
\includegraphics[width=\columnwidth]{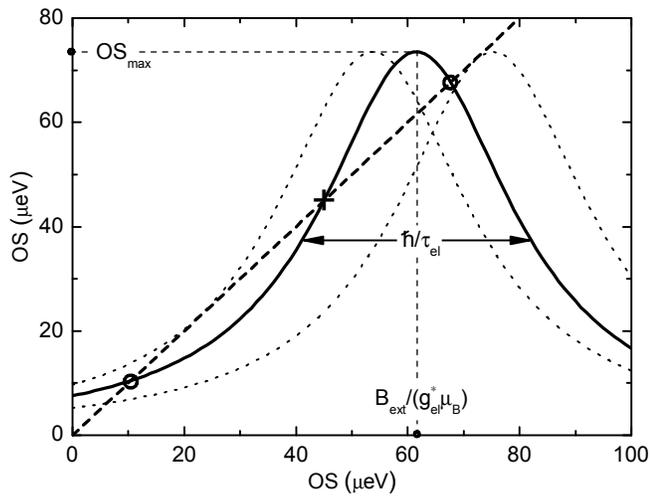}
\caption{\label{FigGraphSol}\textbf{Graphical solution of equation
(\ref{IzSteadySt}):} solid (dashed) line: right (left) hand side
of equation (\ref{IzSteadySt}). These terms correspond to gain and
loss of DNSP, respectively. Circles (cross) indicate the stable
(unstable) solutions for nuclear spin polarization. The center of
the Lorentzian shifts proportionally to the external magnetic
field, explaining the magnetic field dependence of DNSP. The two
dotted curves show the situation at the critical fields $B_1$ and
$B_2$. It can be seen directly from the figure that: 1.)
Bistability can only be observed if the slope of the Lorentzian at
its half width is bigger than 1 and 2.) the difference between the
two critical external fields where either one of the stable
solutions vanishes is on the order of the width of electron spin
states in units of magnetic fields $\hbar/(\tau_{\rm el} g_{\rm
el}^*\mu_{\rm B})$. }
\end{figure}

In this section, we focus on the bistable behavior of the coupled
electron-nuclear spin system in the magnetic field range close to
the ``breakdown'' of DNSP. Figure~\ref{FigGraphSol} shows a
graphical representation of the solutions of the nonlinear
equation (\ref{IzSteadySt}). The result suggests that the maximal
achievable degree of DNSP in our system leads to a maximal OS
given by ${\rm
OS_{max}}=\frac{4}{3}\overline{^{\vphantom{i}}A_i}\,\overline{I^i(I^i+1)}\langle
S_z \rangle(1+T_{\rm 1e}^0/T_{\rm d})^{-1}$. this value is reached
when nuclear spin relaxation is maximized, i.e. when the total
electron Zeeman splitting is zero (cf. equation
(\ref{IzSteadySt})~). It can also be seen from the figure that
there is a regime of external magnetic fields where two stable
solutions for DNSP coexist. One solution leads to a high degree of
nuclear polarization, reaching $\rm{OS_{max}}$ at its maximum,
while the other one shows a low degree of nuclear polarization.
The graphical solution also shows that bistability is an inherent
property of the solutions of equation (\ref{Rateequation}) for
systems where $\rm{OS_{max}}$ is at least on the order of the
width of the electronic spin states ($\hbar/\tau_{\rm el}$), which
is typically the case for localized carriers such as in QDs, but
not for bulk systems. The two stable solutions can be understood
as follows: When increasing an external field while creating a
nuclear field in the opposite direction, the electron Zeeman
splitting is reduced compared to the case of no nuclear
polarization. This keeps nuclear spin relaxation at a high rate
$T_{\rm 1e}^{-1}$ such that DNSP can be maintained. As soon as
$\rm{OS_{max}}$ is reached, however, the system can no further
compensate for an increasing external magnetic field. DNSP will
start to drop, which eventually leads to an abrupt jump of DNSP to
a low value at an external field $B_{1}$. This jump is due to the
negative feedback of the low DNSP on $T_{\rm 1e}^{-1}$. When
ramping the external field down again, now in the absence of
nuclear polarization, the system will initially remain in a state
of low DNSP since $T_{\rm 1e}^{-1}$ is still low. DNSP will
slightly increase though due to the increasing rate $T_{\rm
1e}^{-1}$ of nuclear polarization with decreasing magnetic field
strength. At a field $B_{2}$, the positive feedback of increasing
DNSP on $T_{\rm 1e}^{-1}$ will take over and an abrupt jump to a
state of high nuclear polarization will occur. As can be seen from
figure \ref{FigGraphSol}, the difference between the fields
$B_{1}$ and $B_{2}$ is on the order of the width of the electronic
spin states in units of magnetic fields.

A hint for bistability in the present system can already be seen
in the fit shown in figure~\ref{FigZeemanOSPol}(b). In order to
observe the hysteretic behavior of the system we performed a
magnetic field dependent PL experiment as described above, now by
exciting the QD with light of constant helicity and by ramping the
magnetic field from low to high and back again. Hysteretic
behavior can be expected if the nuclear fields created in that way
are pointing against the external magnetic field. In our system
such a situation is realized when exciting the QD with $\sigma^+$
light and applying an external field in the positive
$z$-direction.

\begin{figure}[t]
\includegraphics[width=\columnwidth]{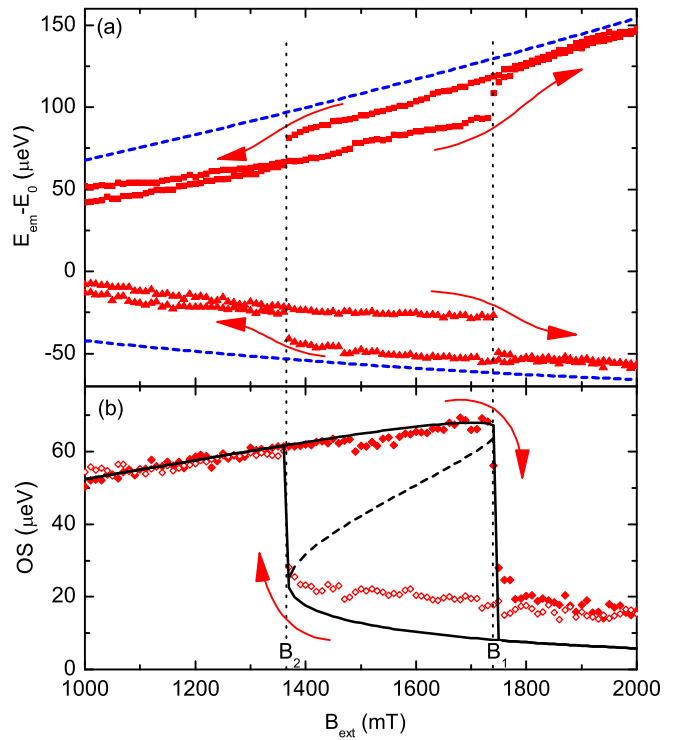}
\caption{\label{FigHysteresis}(color online). \textbf{Hysteresis
behavior of the coupled electron-nuclear spin-system:} Magnetic
field sweeps under excitation with constant light polarization
($\sigma^+$). \textbf{(a)} $X^{-1}$ emission energies, sweeping
magnetic field up or down (indicated by arrows). Squares
(triangles) denote co- (cross-) circular detection with respect to
excitation polarization. The dashed line is a fit to the case of
linearly polarized excitation. \textbf{(b)} Overhauser shifts
extracted from the data shown in (a) for the magnetic field
sweeping up and down (solid and open diamonds, respectively). The
black line shows the simulations described in the text.}
\end{figure}

Figure~\ref{FigHysteresis} shows data obtained in this regime:
Going from low to high field amplitude, DNSP is significant up to
a magnetic field value of $B_{1}=1.74~$T where it suddenly drops.
Sweeping the magnetic field back to low field amplitudes DNSP
reappears at a field $B_{2}=1.36~$T, a value different from
$B_{1}$. The difference of $380~$mT between these two field is on
the order of $\hbar/(\tau_{\rm el} g_{\rm el}^* \mu_{\rm B})$ as
predicted by the model.

Also shown in figure~\ref{FigHysteresis} is a fit of equation
(\ref{EquilibriumI}) to the data. The parameters used for this fit
were
$T_{\rm 1e}^0/T_{\rm d}=4.54$, $\rho_c=0.84$, $\tau_{\rm
el}=32~$ps, $g_{\rm el}^*=-0.68$, which are consistent with the
parameters used in the fit shown in figure~\ref{FigZeemanOSPol}.
As in the previous fit, the reminiscent nuclear polarization at
high fields observed in the experiment is slightly higher than
what is predicted by the model.

\section{Conclusion}

To summarize, we presented a study of the magnetic field
dependence of DNSP in a single resonantly pumped QD. We show clear
evidence of the nonlinear behavior of the tightly coupled electron
and nuclear spin system and show hysteresis as one example of
(spin-)memory of the nuclear spins. A simple rate equation model
is used to reproduce and fit our experimental findings. From the
model and the experimental data we deduce that
a spin polarized electron in an external field can create a
nuclear field that actually overcompensates the applied field: in
this situation, the electrons and holes in the QD feel effective
magnetic fields pointing in opposite directions. Increasing the
external field leads to higher nuclear fields, until a maximal
achievable nuclear polarization is reached at $B_{\rm
ext}=B_1=1.74~$T. At that point, the total effective magnetic
field acting on the electron is zero. This point is of particular
interest because it enables a direct measure of the maximal
nuclear field which we find to be $B_{\rm nuc}=1.74~$T and of the
hole g-factor $g_h^*=-1.2$ .

The experiment along with the model also shows that the maximal
nuclear polarization of $\sim18\%$ achieved in our system is
limited by the fraction $T_{\rm d}/T_{\rm 1e}^0$, i.e. the ratio
between nuclear spin decay time and electron mediated nuclear spin
relaxation time.
While
$T_{\rm d}$ is a fixed parameter given by the nature of the QD,
$T_{\rm 1e}^0$ could potentially be modified by varying the pump
power or the details of the excitation process.\cite{Braun2006}

An extension of the model which includes the dynamics of the
electron spin shows that this spin is linked to the nuclear spins
through a linear relationship. This suggests that not only nuclear
spins but also the electron spin shows a bistable behavior at
certain experimental conditions. This speculation could be further
tested in an experiment on the positively charged exciton where PL
light polarization directly probes the mean electron spin.

The qualitative disagreement of the model with our data in the low
field range where the measured DNSP shows a clear ``kink'' as a
function of magnetic field, indicates that our simple approach
does not give a full description of the nonlinear processes that
lead to an equilibrium state of spins in a QD. A further extension
of the model could include quadrupolar couplings of the nuclei
which would
most effectively depolarize the nuclei at low external fields.
While our rate equation approach was purely classical, it could
also be conceived that the quantum mechanical nature of the
electron spin system would alter the behavior of DNSP at low
fields and explain the unpredicted features in our measurement. To
confirm this hypothesis further theoretical studies would be
required.

\begin{acknowledgments}
We would like to acknowledge O. Krebs for fruitful discussions
about the modelling and helpful inputs about the hysteresis
measurements. Further, we would like to thank V.I. Fal'ko,
G.Burkhard, S.D. Huber and A. H\"ogele for interesting discussions
and help with the manuscript and J. Dreiser for assistance with
sample preparation. This work is supported by NCCR-Nanoscience.
\end{acknowledgments}

\bibliography{BibNuclearSpin}

\end{document}